\documentclass[twoside,twocolumn,9pt]{article}
\usepackage{extsizes}
\usepackage[super,sort&compress,comma]{natbib} 
\usepackage[version=3]{mhchem}
\usepackage[left=1.5cm, right=1.5cm, top=1.785cm, bottom=2.0cm]{geometry}
\usepackage{balance}
\usepackage{times,mathptmx}
\usepackage{sectsty}
\usepackage{graphicx} 
\usepackage{lastpage}
\usepackage[format=plain,justification=justified,singlelinecheck=false,font={stretch=1.125,small,sf},labelfont=bf,labelsep=space]{caption}
\usepackage{hyperref}
\usepackage{float}
\usepackage{fancyhdr}
\usepackage{fnpos}
\usepackage[english]{babel}
\usepackage[utf8]{inputenc} % usually not needed (loaded by default)
\usepackage[T1]{fontenc}
\addto{\captionsenglish}{%
  
}
\usepackage{array}
\usepackage{droidsans}
\usepackage{charter}
\usepackage[T1]{fontenc}
\usepackage[usenames,dvipsnames]{xcolor}
\usepackage{setspace}
\usepackage[compact]{titlesec}
\usepackage{hyperref}
\usepackage{epstopdf}%This line makes .eps figures into .pdf - please comment out if not required.

\definecolor{cream}{RGB}{222,217,201}

\begin{document}

\pagestyle{fancy}
\thispagestyle{plain}
\fancypagestyle{plain}{

%%%HEADER%%%
\fancyhead[C]{\includegraphics[width=18.5cm]{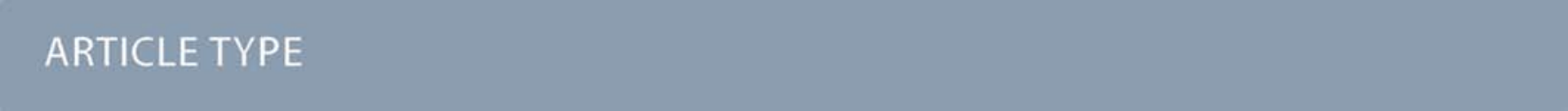}}
\fancyhead[L]{\hspace{0cm}\vspace{1.5cm}\includegraphics[height=30pt]{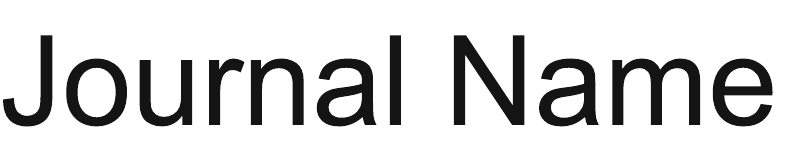}}
\fancyhead[R]{\hspace{0cm}\vspace{1.7cm}\includegraphics[height=55pt]{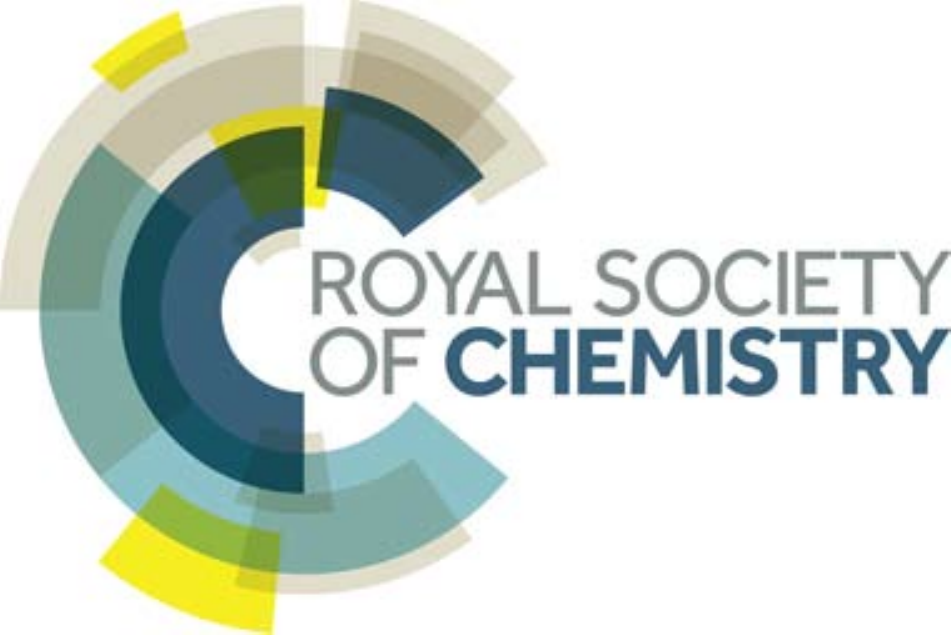}}
\renewcommand{\headrulewidth}{0pt}
}
%%%END OF HEADER%%%

%%%PAGE SETUP - Please do not change any commands within this section%%%
\makeFNbottom
\makeatletter
\renewcommand\LARGE{\@setfontsize\LARGE{15pt}{17}}
\renewcommand\Large{\@setfontsize\Large{12pt}{14}}
\renewcommand\large{\@setfontsize\large{10pt}{12}}
\renewcommand\footnotesize{\@setfontsize\footnotesize{7pt}{10}}
\makeatother

\renewcommand{\thefootnote}{\fnsymbol{footnote}}
\renewcommand\footnoterule{\vspace*{1pt}% 
\color{cream}\hrule width 3.5in height 0.4pt \color{black}\vspace*{5pt}} 
\setcounter{secnumdepth}{5}

\makeatletter 
\renewcommand\@biblabel[1]{#1}            
\renewcommand\@makefntext[1]% 
{\noindent\makebox[0pt][r]{\@thefnmark\,}#1}
\makeatother 
\renewcommand{\figurename}{\small{Fig.}~}
\sectionfont{\sffamily\Large}
\subsectionfont{\normalsize}
\subsubsectionfont{\bf}
\setstretch{1.125} %In particular, please do not alter this line.
\setlength{\skip\footins}{0.8cm}
\setlength{\footnotesep}{0.25cm}
\setlength{\jot}{10pt}
\titlespacing*{\section}{0pt}{4pt}{4pt}
\titlespacing*{\subsection}{0pt}{15pt}{1pt}
%%%END OF PAGE SETUP%%%

%%%FOOTER%%%
\fancyfoot{}
\fancyfoot[LO,RE]{\vspace{-7.1pt}\includegraphics[height=9pt]{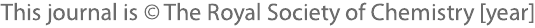}}
\fancyfoot[CO]{\vspace{-7.1pt}\hspace{13.2cm}\includegraphics{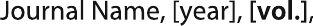}}
\fancyfoot[CE]{\vspace{-7.2pt}\hspace{-14.2cm}\includegraphics{RF}}
\fancyfoot[RO]{\footnotesize{\sffamily{1--\pageref{LastPage} ~\textbar  \hspace{2pt}\thepage}}}
\fancyfoot[LE]{\footnotesize{\sffamily{\thepage~\textbar\hspace{3.45cm} 1--\pageref{LastPage}}}}
\fancyhead{}
\renewcommand{\headrulewidth}{0pt} 
\renewcommand{\footrulewidth}{0pt}
\setlength{\arrayrulewidth}{1pt}
\setlength{\columnsep}{6.5mm}
\setlength\bibsep{1pt}
%%%END OF FOOTER%%%

%%%FIGURE SETUP - please do not change any commands within this section%%%
\makeatletter 
\newlength{\figrulesep} 
\setlength{\figrulesep}{0.5\textfloatsep} 

\newcommand{\topfigrule}{\vspace*{-1pt}% 
\noindent{\color{cream}\rule[-\figrulesep]{\columnwidth}{1.5pt}} }

\newcommand{\botfigrule}{\vspace*{-2pt}% 
\noindent{\color{cream}\rule[\figrulesep]{\columnwidth}{1.5pt}} }

\newcommand{\dblfigrule}{\vspace*{-1pt}% 
\noindent{\color{cream}\rule[-\figrulesep]{\textwidth}{1.5pt}} }

\makeatother
%%%END OF FIGURE SETUP%%%

%%%TITLE, AUTHORS AND ABSTRACT%%%
\twocolumn[
  \begin{@twocolumnfalse}
\vspace{3cm}
\sffamily
\begin{tabular}{m{4.5cm} p{13.5cm} }

\includegraphics{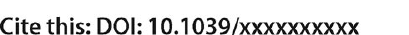} & \noindent\LARGE{\textbf{Behaviour of flexible superhydrophobic striped surfaces during (electro-)wetting of a sessile drop $\dag$}} \\%Article title goes here instead of the text "This is the title"
\vspace{0.3cm} & \vspace{0.3cm} \\

 & \noindent\large{Arvind Arun Dev,\textit{$^{a,c}$} Ranabir Dey,\textit{$^{b,c}$} and Frieder Mugele\textit{$^{c*}$}} \\%Author names go here instead of "Full name", etc.

\includegraphics{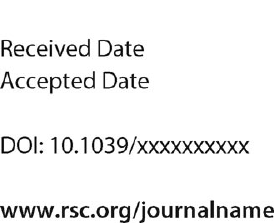} & \noindent\normalsize{}  We study here the microscopic deformations of elastic lamellae constituting a superhydrophobic substrate under different wetting conditions of a sessile droplet using electrowetting. The deformation profiles of the lamellae are experimentally evaluated using confocal microscopy. These experimental results are then explained using a variational principle formalism within the framework of linear elasticity. We show that the local deformation profile of a lamella is mainly controlled by the net horizontal component of the capillary forces acting on its top due to the pinned droplet contact line. We also discuss the indirect role of electrowetting in dictating the deformation characteristics of the elastic lamellae. One important conclusion is that the small deflection assumption, which is frequently used in the literature, fails to provide a quantitative description of the experimental results; a full solution of the non-linear governing equation is necessary to describe the experimentally obtained deflection profiles.\\   

\end{tabular}

 \end{@twocolumnfalse} \vspace{0.6cm}
]
%%%END OF TITLE, AUTHORS AND ABSTRACT%%%

%%%FONT SETUP - please do not change any commands within this section
\renewcommand*\rmdefault{bch}\normalfont\upshape
\rmfamily
\section*{}
\vspace{-1cm}

%%%FOOTNOTES%%%

\footnotetext{\textit{$^{a}$~Current affiliation: Department of magnetism of nanostructured objects (DMONS), Institut de Physique et Chimie des
Mat{\`e}riaux de Strasbourg (IPCMS), CNRS UMR 7504,  Universit{\`e} de Strasbourg, 23 Rue du Loess, 67034 Strasbourg, France.\\
$^{b}$~ Current affiliation: Dynamics of Complex Fluids, Max Planck Institute for Dynamics and Self-organization, Am Fassberg 17, 37077 Goettingen, Germany.\\
$^{c}$~Physics of Complex Fluids, MESA+ Institute for Nanotechnology, University of Twente, PO Box 217, 7500 AE Enschede, The Netherlands}}

%Please use \dag to cite the ESI in the main text of the article.
%If you article does not have ESI please remove the the \dag symbol from the title and the footnotetext below.
\footnotetext{\dag~Electronic Supplementary Information (ESI) available: [details of any supplementary information available should be included here]. See DOI: 10.1039/b000000x/}
%additional addresses can be cited as above using the lower-case letters, c, d, e... If all authors are from the same address, no letter is required

\footnotetext{$^*$~ e-mail: f.mugele@utwente.nl}

%%%END OF FOOTNOTES%%%

%%%MAIN TEXT%%%%

\section{Introduction}
Superhydrophobic surfaces are ubiquitous in nature like lotus leaves\cite{Art_LotusLeaf} and salvinia surfaces  \cite{Art_SalviniaSurface}. While a lotus leaf surface can be considered as a rigid superhydrophobic surface, a Salvinia surface is covered with hairy hierarchical structures with elastic properties and may be considered as a flexible superhydrophobic surface \cite{Otten}. Artificial surfaces mimicking the hierarchical structures and extreme water repellency of such natural superhydrophobic surfaces are intrinsic to various technologies like anti-freezing surfaces\cite{Art_3}, self-cleaning surfaces\cite{Art_4}, corrosion resistance surfaces\cite{Art_5}, enhanced water vapor condensation\cite{Art_6} and fog harvesting\cite{Art_7}. Furthermore, over the years artificial superhydrophobic surfaces have gained importance in biomedical applications such as control of protein adsorption \cite{Art_8}, cellular interaction \cite{Art_9}, anti-platelet adhesion \cite{Art_10}, and reduction of bacterial growth\cite{Art_11}. Specifically, flexible structured surfaces are used for quantifying cell traction forces \cite{Art_CellTrac1,Art_CellTrac2}, and as flow sensors \cite{Art_ArtiFlowSens}, artificial skin\cite{Art_ArtiSkin1}, and tactile sensors\cite{Art_TactileSensor}.

Owing to such a wide range of technological applications, extensive research has been performed to understand the wetting characteristics of liquids on both natural and artificial superhydrophobic surfaces \cite{Art_Advancing,Art_Transition,Art_14,Art_15, QuereReview, XiaAnisoReview}. However, these studies focus on the wetting behaviour on rigid structured substrates; studies pertaining to wetting characteristics on substrates with soft/flexible structures are relatively scarce despite the wide range of applications. The existing studies in this regard reveal that the wetting characteristics on soft micro-structured substrates are definitely dependent on the elasticity of the underlying structures. The droplet rolling contact angle increases, and the receding contact angle decreases, with increasing softness of the elastic micro-structures \cite{Art_ContElas}. Furthermore, contact angle hysteresis often increases due to the deformation and collapse of the soft micro-structures underneath the drop \cite{Art_ContAng}. However, these studies do not examine in detail the microscopic deformation of the underlying soft structures due to capillary interactions during droplet wetting, specifically when the droplet is still in the Cassie state \cite{CassieState,Art_Advancing}. In order to have a comprehensive idea of wetting and superhydrophobicity on soft structured substrates, it is imperative to understand the changes in shape of the underlying structures due to the involved capillary interactions. 

The deformation of an array of soft micropillars, constituting a superhydrophobic substrate, during wetting was studied very recently \cite{Art_BatEl}. This work highlights the relationship between the deformation of the underlying soft micropillars and the macroscopic wetting characteristics of a glycerol/water drop as characterized by the apparent advancing and receding contact angles \cite{Art_BatEl}. However, the behaviour of the soft micropillars was analyzed assuming small deflections of the elastic pillars, which may not be the general case considering the wide range of applications involving different magnitudes of capillary forces and structure geometries. Therefore, a detailed analysis of the deformation characteristics of flexible superhydrophobic substrates over a wide range of wetting conditions, and hence, capillary force magnitudes, has remained unexplored.  

In this paper we investigate the deformation of flexible stripes/lamellae, constituting a superhydrophobic substrate, under different wetting conditions of a sessile droplet using electrowetting (EW) \cite{mugele2019electrowetting}. During EW, the presence of surface charges results in a Maxwell stress distribution at the droplet liquid-vapour interface, which gives rise to a net electrical force pulling on it \cite{Manukyan,Oh_Mugele}. This electrical force results in enhanced wetting and deformation of the drop on a macroscopic scale. Moreover, at equilibrium the balance of the Maxwell stress and the Laplace pressure results in large curvature of the liquid-vapour interfaces. So, during EW on a structured superhydrophobic substrate, provided the droplet is still in Cassie state, the enhanced wetting, macroscopic deformation of the drop, and the increasing curvature of the liquid-vapour interface in between the structures can result in the application of different magnitudes of capillary forces on the structures at the droplet contact line. Therefore, EW can provide a flexible way for probing the mechanics of the soft deformable structures over a range of capillary force magnitude. This is precisely what we try to exploit in the present work. Additionally, the present work also explores the feasibility of using such soft superhydrophobic substrates for the myriad of EW applications \cite{Wheeler,Shamai}. We perform a systematic study of the deformation shapes of the soft lamellae during EW of a sessile drop for different structure geometry, material elasticity, and EW conditions using confocal microscopy. We also explain the observed deformation characteristics using the classical elastica theory which goes beyond the small deflection assumption.

\section{Materials and Methods}
\subsection{Fabrication of striped superhydrophobic substrates of different elasticity}
\begin{figure}[!htbp]
\centering
\includegraphics[width=9cm, height=7.4cm]{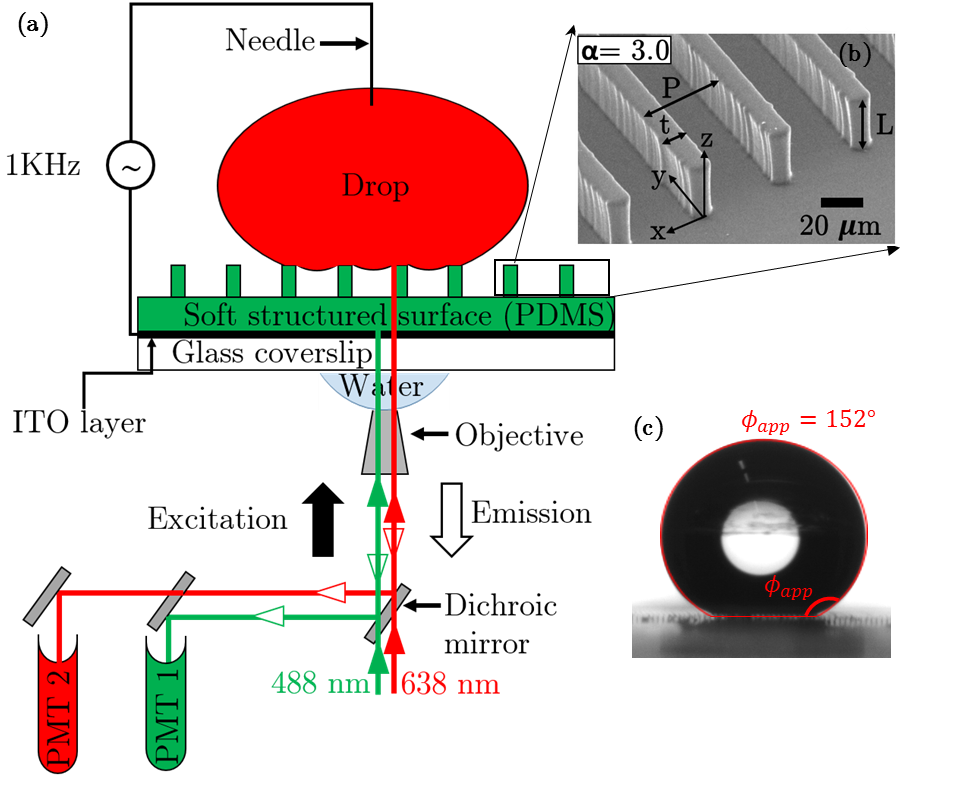}
\caption{\label{fig_1} (a) Schematic of the experimental setup used for visualizing the shapes of the flexible lamellae during EW of a sessile drop. (b) A SEM image of a soft superhydrophobic striped surface fabricated using a three-step soft-molding process. (c) A water droplet at equilibrium on the soft superhydrophobic striped surface.}
\end{figure}

We use Sylgard $^ \textup{TM}$ 184, a two part, thermo-curable, cross-linked polydimethylsiloxane (PDMS) based elastomer to fabricate soft striped superhydrophobic surfaces of varying rigidity. The soft structured substrates are fabricated using a three step soft molding process \cite{Art_BatEl}. In the first step, we fabricate a SU8 replica (primary) of the desired striped substrate using photo-lithography (see step 1 in Sec. S1 in ESI). In the second step, we prepare a mold out of the SU8 replica using Sylgard 184 prepolymer. The prepolymer is prepared by mixing the base and the cross-linker in the weight ratio of 10:1. This mixture is then poured on the primary SU8 replica, and cured in vacuum at 60$^ \circ$C for 14 hrs. The mold is obtained by peeling off the cured Sylgard 184 from the SU8 replica (see step 2 in Sec. S1 in ESI). The secondary mold is further cleaned using air plasma for 45 seconds at 60 W power, and thereafter silanized with trichloro (1H,1H,2H,2H-perfluoro-octyl) silane (Sigma-Aldrich) using chemical vapour deposition. In the third step, Sylgard 184 is first prepared by mixing the base and the cross-linker in different weight ratios in order to fabricate substrates of different rigidity, as quantified by the Young's modulus of elasticity ($E$). For the present work, we use three different base to crosslinker ratios- 10:1, 15:1, and 20:1, resulting in values of $E$ varying from 2.1 MPa to 0.6 MPa (for details of the measurement of $E$ see sub-section 2.2). Thereafter, an ITO coated coverslip of thickness 175 $\mu$m  is spin coated with the Sylgard 184 prepolymer (mixed in the desired weight ratio as mentioned above). The Sylgard 184 coated coverslip is then placed on top of the secondary Sylgard 184 mold, and is then cured for 14 hrs at 60$^ \circ$C. After curing, the secondary mold is peeled off from the coverslip, which leaves behind the final replica of the soft striped substrate on the ITO coverslip (Fig. \ref{fig_1} (b); also see step 3 in Sec. S1 in ESI). The co-ordinate system shown in Fig. \ref{fig_1}(b) is followed throughout the subsequent analysis. The length of each lamella along the y-axis is 3 mm, while the distance between any two lamellae is 37 $\mu$m. The other dimensions of the lamellae constituting the striped substrates of different elasticity are listed in Table 1 (and shown in Fig. \ref{fig_1}(b)).

\begin{table}[!h]
\small
  \caption{\ Measured dimensions of lamellae}
  \label{tbl:example}
  \begin{tabular*}{0.48\textwidth}{@{\extracolsep{\fill}}llll}
    \hline
    Sample No. & Height (L) ($\mu m$) & Width (t) ($\mu m$) & Aspect Ratio ($\alpha=L/t$)  \\
    \hline
    \multicolumn{4}{c}{Sylgard 10:1 ($E=2.1$ MPa)}\\
    \hline
    1 & 34 & 11.3 & 3.0 \\
    2 & 34 & 15 & 2.2 \\
    3 & 34 & 25 & 1.3 \\
    \hline
    \multicolumn{4}{c}{Sylgard 15:1 ($E=0.9$ MPa)}\\
    \hline
    4 & 30.75 & 13.9 & 2.2 \\
    5 & 30.75 & 24 & 1.3\\
   \hline
    \multicolumn{4}{c}{Sylgard 20:1 ($E=0.6$ MPa)}\\
    \hline
    6 & 30.75 & 13.9 & 2.2 \\
    7 & 30.75 & 24 & 1.3\\
    \hline
  \end{tabular*}
\end{table}
\subsection{Measurement of Young's Modulus}
The values of the Young's modulus $(E)$ of cured Sylgard 184 substrates prepared by mixing the base and the cross-linker in the weight ratios of 10:1, 15:1, and 20:1 are measured independently using a extensometer (Zwick Roell). The extensometer provides stress-strain relations for the different Sylgard 184 substrates, which remain linear for small strains. $E$ is evaluated from the slope of the stress strain curve in the linear region, in accordance with the Hooke's law \cite{Art_31}. The values of $E$ for base to crosslinker ratios of 10:1, 15:1, and 20:1 are 2.1 MPa, 0.9 MPa, and 0.6 MPa respectively (see Sec. S2 in ESI for details). 

\subsection{Experimental setup}
The experimental setup used for visualising the behaviour of the flexible striped superhydrophobic substrates during EW of a sessile drop is shown in Fig. \ref{fig_1}(a). For ac-EW, a 4 $\mu$l drop of 1 mM aqueous potassium chloride (KCl) solution is first placed on the substrate (Fig. \ref{fig_1}(a), (c)), and then the desired electrical voltage is applied between the ITO electrode (on the coverslip) and a needle electrode dipped into the droplet ((Fig. \ref{fig_1}(a)). The electrical voltage is applied by means of a function generator and an amplifier. For the experiments reported here, the root-mean-square voltage $(U_{rms})$ is varied from 0V to 450V, while the applied frequency is kept fixed at $f=1$ kHz. The deformation of the flexible lamellae due to interfacial capillary interactions during EW is characterised using confocal microscopy. 
\begin{figure}[!h]  
\centering
\includegraphics[width=5.88cm, height=14cm]{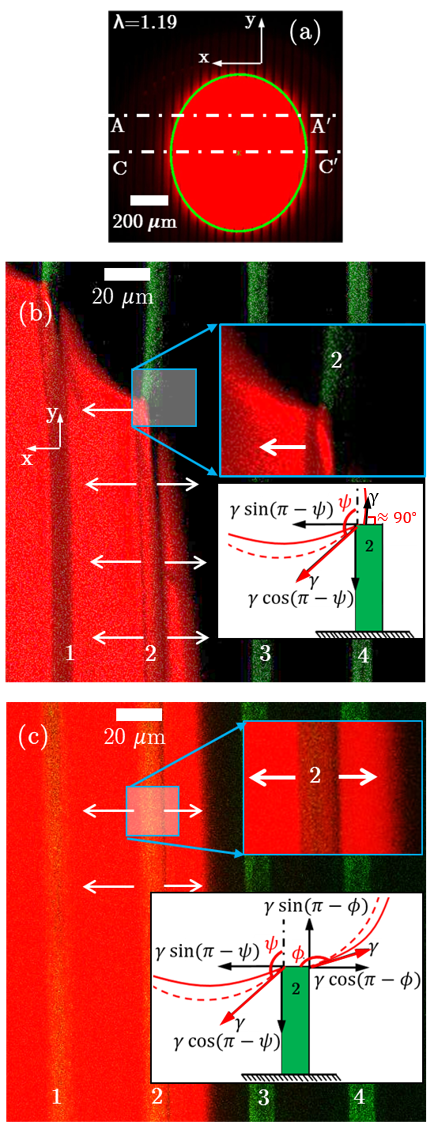}
\caption{\label{fig_2}(a) Footprint of a sessile drop on the soft supehydrophobic striped surface captured using a confocal microscope. The ellipticity of the droplet footprint is measured by the ratio ($\lambda$) of the major axis (along the lamellae) to the minor axis (perpendicular to the lamellae). xy projections of the two vertical planes along which the xz confocal slices are acquired are shown by A-A$^\prime$ and C-C$^\prime$. (b) Close-up of the macroscopic droplet contact line in the xy plane about A-A$^\prime$. The image clearly shows the deformation of lamella 2 due to the particular orientation of the contact line such that it covers only one top edge (left top edge) of the lamella; the blow-up in the inset clearly shows the partial coverage of the top of lamella 2 where it is deformed. (c) Close-up of the macroscopic droplet contact line in the x-y plane about C-C$^\prime$. In this case, the droplet covers both the top edges of the outermost lamella underneath the droplet, i.e. lamella 2, as also shown by the blow-up in the inset. The schematics in the insets in (b) and (c) show the capillary forces acting on the top edges of the lamella due to the pinned three phase contact lines at A-A$^\prime$ and C-C$^\prime$ respectively.}
\end{figure}

\begin{figure*}[!h]
\centering
\includegraphics[width=18cm, height=13.63cm]{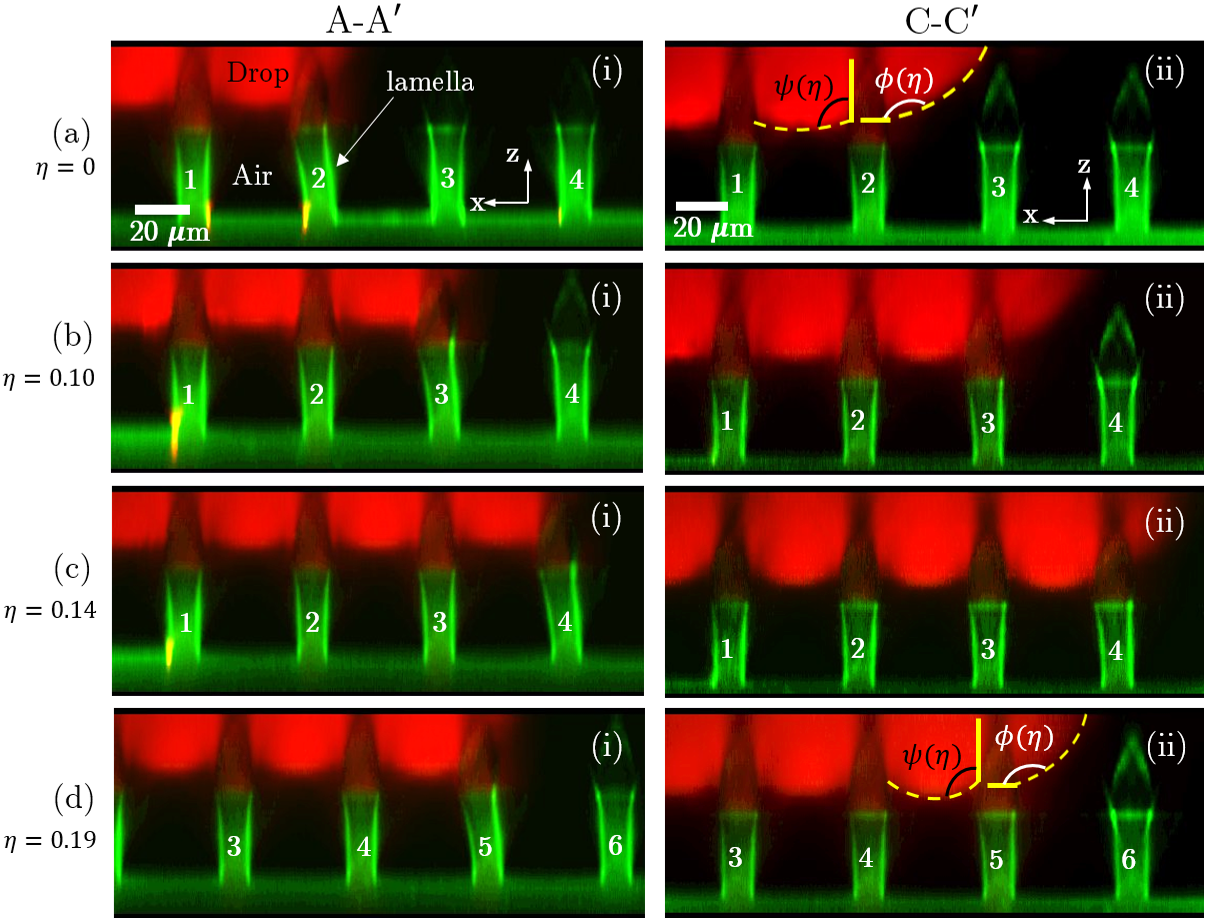}
\caption{\label{fig_3} xz confocal slices showing a droplet (red) in Cassie state on the soft structured surface (green) at the A-A$^\prime$ plane ((i) left column) and at the C-C$^\prime$ plane ((ii) right column) for increasing magnitude of the applied voltage (rows), as represented by the increasing value of the electrowetting number $\eta$-- (a) $\eta=0$, (b) $\eta=0.10$, (c) $\eta=0.14$, and (d) $\eta=0.19$. As the droplet spreads with increasing $\eta$, the droplet contact line jumps from one lamella to the next, while maintaining the Cassie state. At a particular value of $\eta$, the deflection of the lamella at the macroscopic droplet contact line (`2' in (a), `3' in (b), `4' in (c), `5' in (d)) is clearly visible, especially at A-A$^\prime$.  $\psi(\eta)$ denotes the local contact angle at which the liquid-air interface meets the top inner edge of the deflected lamella underneath the droplet, and $\phi(\eta)$ is the angle at which the macroscopic droplet contact line meets the lamella edge. The aspect ratio $(\alpha)$ and the Young`s modulus $(E)$ of the lamellae shown here are 3.0 and 2.1 MPa respectively.}
\end{figure*}

\subsection{Details of confocal microscopy}
For confocal microscopy we use two fluoroscent dyes-- DFSB-K175 (Risk Reactor) in the Sylgard 184 substrate and Alexa Fluor$^ \textup{TM}$ 647 (ThermoFisher SCIENTIFIC) in the drop. The dye DFSB-K175 is mixed with the crosslinker of Sylgard 184 in the volume ratio of 1:3000 during the coating of the ITO cover slip in the third step of the substrate fabrication procedure. Alexa Fluor 647 is first mixed with Milli-Q water in the concentration of 5 $\mu$g/ml to create a stock solution; this stock solution is further diluted with 1mM KCl solution in the volume ratio of 1:100 and used for the confocal experiments. The maximum emission wavelength for the dye DFSB-K175 corresponding to an excitation wavelength of 488 nm is 540 nm, while the maximum excitation and emission wavelengths for Alexa Fluor 647 are 653 nm and 669 nm respectively (see Sec. S3 in ESI for details).

For confocal imaging (Fig. \ref{fig_1}(a)) we use a Nikon A1 inverted line scanning confocal microscope with excitation lasers at 488 nm (for DFSB-K175 in the substrate) and 638 nm (for Alexa Fluor 647 in the drop), and two objectives-- 60x water immersion with numerical aperture (NA)=1.2 and 20x dry with NA=0.85. The emissions from the two dyes are collected using band filters in the range 500 nm to 550 nm for DFSB-K175 and in the range of 663 nm to 700 nm for Alexa Fluor 647. The refractive indices of the different components involved are as follows-- glass: 1.5, Sylgard 184: 1.42, air: 1.0, and water drop: 1.33. 3D confocal scans are performed in the immediate vicinity of the macroscopic droplet contact line such that lamellae wetted by the droplet as well as outside the droplet footprint are simultaneously visible (Fig. \ref{fig_2}). The confocal scans are performed at two different locations (Fig. \ref{fig_2}(a)) so that two different orientations of the macroscopic droplet contact line relative to a lamella can be investigated (Fig. \ref{fig_2}(b) and (c)). xy projections of the two vertical planes along which xz slices are acquired for studying the deflection profiles of the lamella are schematically represented by A-A$^\prime$ and C-C$^\prime$ in Fig. \ref{fig_2}(a); note that the planes for the xz slices are always perpendicular to the lamellae. Furthermore, C-C$^\prime$ passes through the centre of the droplet footprint.  The 3D confocal scans are post-processed using ImageJ, and the xz slices are analyzed using an in-house MATLAB code to evaluate the deformation characteristics of the soft lamellae due to interfacial capillary interactions (see Sec. S4 in ESI for image analysis details).

\begin{figure*}[!htbp]  
\centering
\includegraphics[width=18cm, height=10.13cm]{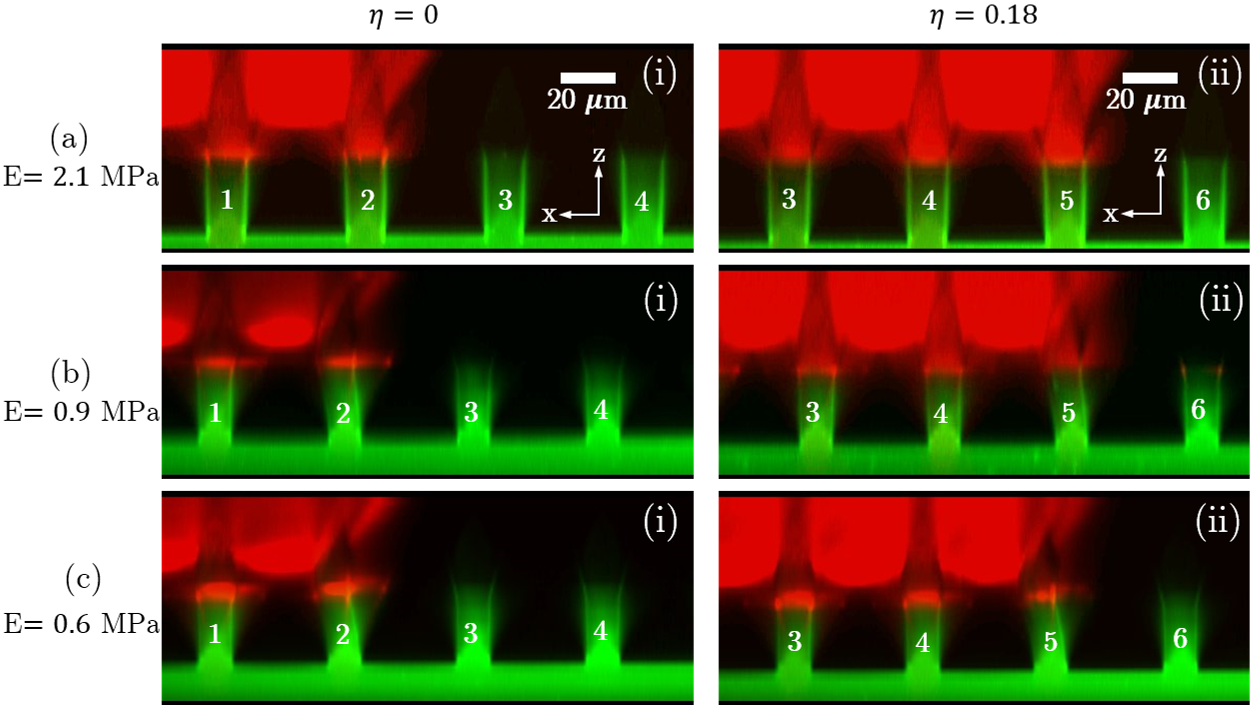}
\caption{\label{fig_4} Increasing deflection of the lamella at the droplet contact line with decreasing value of the Young`s modulus $E$ (rows)-- (a) $E=2.1$ MPa, (b) $E=0.9$ MPa, and (c) $E=0.6$ MPa, for electrowetting number $\eta=0$ ((i) left column) and $\eta=0.18$ ((ii) right column), at A-A$^\prime$. The aspect ratio $(\alpha)$ of the lamellae shown here is 2.2.}
\end{figure*}

\section{Result and discussions}
We investigate the behaviour of the flexible lamellae underneath the electrowetted droplet both at the macroscopic droplet contact line (edge of the droplet) as well as away from the macroscopic droplet contact line. For the deformation of a lamella at the macroscopic droplet contact line we investigate two possible configurations of the contact line relative to the lamella top. One, the configuration in which the macroscopic droplet contact line locally covers only one top edge of a lamella, which is usually the case where the contact line changes orientation and leaves a lamella (see the close-up of the macroscopic droplet contact line on top of lamella 2 at A-A$^\prime$ in Fig. \ref{fig_2}(b)). Two, the configuration in which the macroscopic droplet contact line locally covers both the top edges of a lamella, e.g. in the vicinity of the minor axis (diameter) of the elliptical (circular) droplet footprint perpendicular to the lamellae (see the close-up of the macroscopic droplet contact line on top of lamella 2 at C-C$^\prime$ in Fig. \ref{fig_2}(c)). It must be noted here that the droplet footprint is initially elliptical in shape due to the inherent anisotropy in the substrate, but gradually becomes circular due to spreading under the applied electrical voltage  (see Sec. S5 in ESI). 

In Fig. \ref{fig_3}, the image columns A-A$^\prime$ and C-C$^\prime$ show the configurations of the lamellae in the x-z planes A-A$^\prime$ and C-C$^\prime$ respectively for increasing magnitude (different rows) of the applied electrical voltage represented here by the so-called non-dimensional electrowetting number $\eta$ \cite{mugele2019electrowetting}.Note first that for the range of electrical voltage applied here the droplet progressively undergoes enhanced wetting while always maintaining the Cassie state (see the gradual progression of the macroscopic droplet contact line over the lamellae 2 to 5 with increasing $\eta$ in Fig. \ref{fig_3}(a) to (d) at both A-A$^\prime$ and C-C$^\prime$). Second, during EW only the lamella underneath the macroscopic droplet contact line deforms, while the other lamellae underneath the droplet remain undeformed (e.g. see lamellae 2, 3, 4, and 5 in rows (a), (b), (c), and (d) respectively in Fig. \ref{fig_3}). Third, apparently the magnitude of deflection of the outermost lamella at A-A$^\prime$ is more compared to that observed in C-C$^\prime$ (compare the shapes of the lamella underneath the macroscopic droplet contact line in A-A$^\prime$ and C-C$^\prime$ corresponding to any value of $\eta$ in Fig. \ref{fig_3}). Finally, the curvatures of the liquid-air interfaces in-between the lamellae underneath the droplet increase with increasing $\eta$ due to the increasing Maxwell stress.  Note that the change in curvature of the liquid-air interface between the lamella is weakly dependent on $\eta$ and a clear detection of the same is somewhat restricted by the optical challenges of diffraction. However, comparison of the interfaces as outlined by the yellow dotted line (guide to the eye) in rows (a) and (d) in Fig. \ref{fig_3} makes the increased curvature due to the applied voltage ($\eta=0.19$) still visible. Here, $\psi (\eta)$ denotes the local angle at which the liquid-air interface meets the lamella edge underneath the droplet, while $\phi(\eta)$ denotes the apparent angle the droplet-air interface makes with the horizontal at the macroscopic droplet contact line (Fig. \ref{fig_3}).

\begin{figure*} [!h]
\centering
\includegraphics[width=18cm, height=12.61cm]{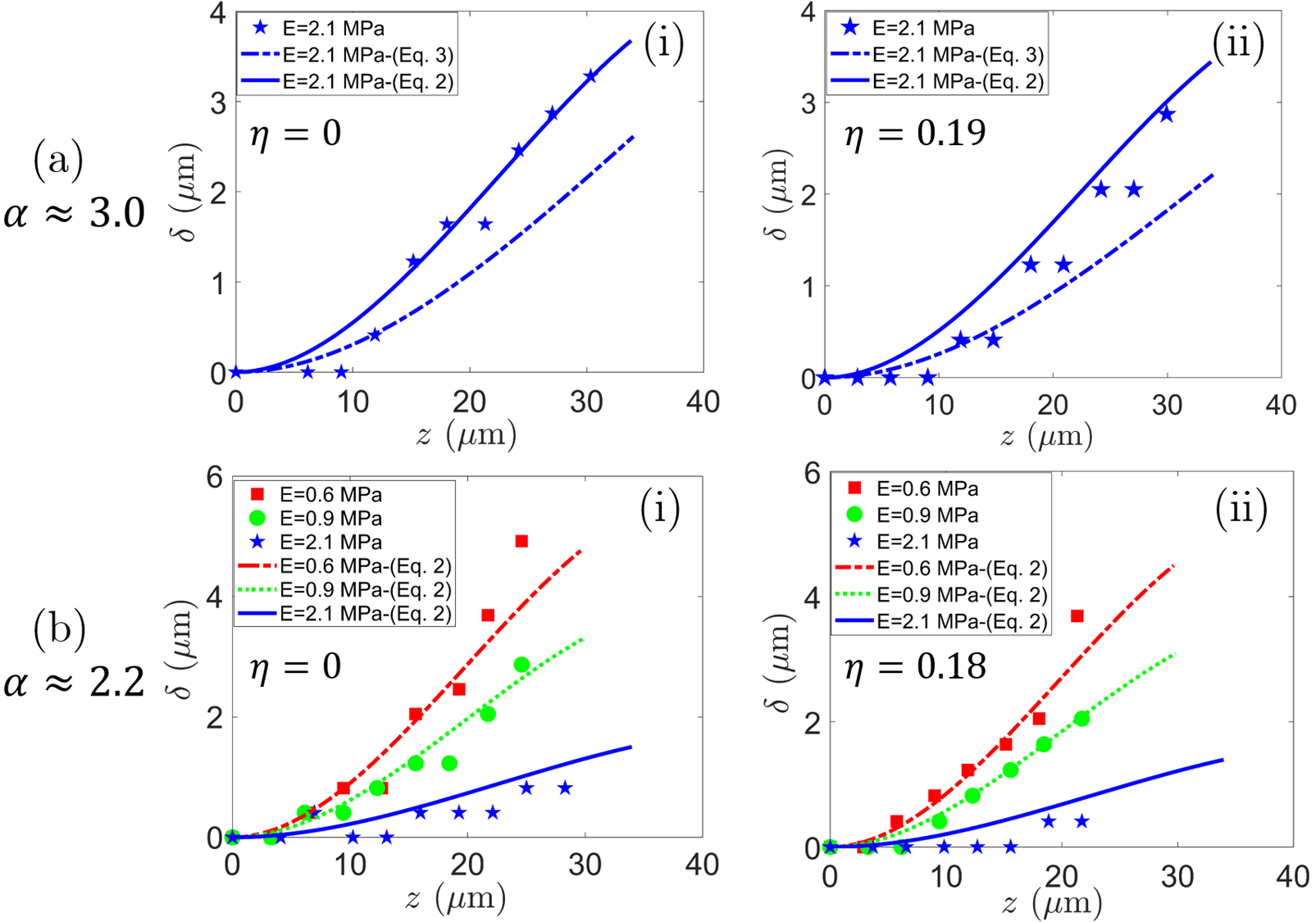}
\caption{\label{fig_5} Variations in the shape of the deflected lamella (in the x-z plane) at the droplet contact line (markers: experimental data; lines: theory) with reducing value of the Young`s modulus $E$, for different values of the aspect ratio $\alpha$ (rows)-- (a) $\alpha=3.0$, (b) $\alpha=2.2$, at two different magnitudes of the electrowetting number $\eta$ ((i) left column-- $\eta=0$; (ii) right column-- $\eta=0.18-0.19$). $\delta(z)$ denotes the deflection of the lamella in the x-z plane at a height $z$ from the base of the lamella. The shapes of the deflected lamella shown here all correspond to the A-A$^\prime$ plane.}
\end{figure*}

In Fig. \ref{fig_4}, the left and right coloumns show the lamellae  configurations for $\eta=0$ and $\eta=0.18$ respectively for decreasing E (different rows). Note that all the lamellae configurations in Fig. \ref{fig_4} are at A-A$^\prime$ and the lamellae have identical aspect ratio $\alpha=2.2$. The deformation of the lamella at the macroscopic droplet contact line (i.e lamella 2 for $\eta=0$ and lamella 5 for $\eta=0.18$) increases with decreasing value of $E$ (Fig. \ref{fig_4}) as qualitatively anticipated. Moreover, a comparison of Fig. \ref{fig_4}(a)(i) and Fig. \ref{fig_3}(a)(i) or Fig. \ref{fig_4}(a)(ii) and Fig. \ref{fig_3}(d)(i) shows that the deformation of the lamella decreases with decreasing $\alpha$ for a particular value of $E$. Furthermore, for a particular value of $E$, there is no significant difference in the deformation of the lamella underneath the macroscopic contact line for different values of $\eta$ (compare images in columns (i) and (ii) corresponding to any row in Fig. \ref{fig_4}). Similar observations were also recorded for soft substrates with lamellae of $\alpha=1.3$ (see Sec. S6 in ESI).

In order to understand the EW-induced deformation characteristics of the soft lamellae for different values of  $\alpha$, $E$, and $\eta$, it is useful to first delineate the orientations of the capillary forces acting on a lamella due to the pinned contact lines. The orientations of the capillary forces acting at the top of the lamella underneath the macroscopic droplet contact line at  A-A$^\prime$ and  C-C$^\prime$ are schematically shown in the insets in Fig. \ref{fig_2}(b) and \ref{fig_2}(c) respectively. The red solid and dashed lines in the schematics represent the water-air interfaces without any applied voltage and under an applied electrical voltage respectively. The curvatures of these interfaces determine the angles $\psi(\eta)$ and $\phi(\eta)$, and consequently, the magnitudes of the horizontal and the vertical components of the capillary forces acting on the top of the lamella (black arrows in the schematics shown in Fig. \ref{fig_2}(b) and Fig. \ref{fig_2}(c)). Note that at A-A$^\prime$, the macroscopic droplet contact line crosses the lamella somewhere between the two top edges of the lamella and the droplet covers only one top edge. Moreover, the apparent contact angle for such a configuration of the macroscopic droplet contact line is always close to 90$^\circ$. Accordingly at A-A$^\prime$, there is an almost vertical force $\gamma$ acting on the top of lamella besides the capillary force acting at the edge of the lamella covered by the drop (see Fig. \ref{fig_2}(b) and left coloumn of Fig. \ref{fig_3}). Furthermore, with increasing $\eta$, $\psi$ increases due to Maxwell stress induced enhanced curvature of the liquid-air interface underneath the droplet at  A-A$^\prime$ and  C-C$^\prime$; however, simultaneously $\phi$ decreases at C-C$^\prime$ due to enhanced wetting with increasing $\eta$. The variation in $\phi$ between A-A$^\prime$ and  C-C$^\prime$ for a given value of $\eta$ is due to the anisotropy in the substrate, which makes the apparent contact angle dependent on the location of the macroscopic droplet contact line on the substrate \cite{AnisotropicWetting}. The overall $\eta$-dependence of $\psi$ and $\phi$ makes the orientation of the capillary forces acting on the outermost lamella at the pinned droplet contact lines function of $\eta$, which at least theoretically makes the consequential deformation of the lamella dependent on $\eta$. So, at A-A$^\prime$ we consider two capillary forces acting on the outermost lamella due to the pinning of the macroscopic droplet contact line-- one acting somewhere between the two top edges of the lamella and the second acting on one edge of the lamella covered by the droplet (left edge in accordance with Fig. \ref{fig_2}(b)); at C-C$^\prime$, we consider the capillary forces on both the top edges of the outermost lamella (both left and right edges in accordance with Fig. \ref{fig_2}(c)). The different components of these capillary forces can be written as following
\begin{subequations}\label{Eqn_1}
    \begin{align}
        F_{xl}=\gamma \sin (\pi-\psi(\eta))   \label{Eqn_1a}\\
         F_{zl}=\gamma \cos (\pi-\psi(\eta)) \label{Eqn_1b}\\
F_{zt}=\gamma \hspace{2cm} \label{Eqn_1b}\\
         F_{xr}=\gamma \cos (\pi-\phi(\eta))  \label{Eqn_1c}\\
        F_{zr}=\gamma \sin(\pi-\phi(\eta)) \label{Eqn_1d}
    \end{align}
\end{subequations}

Here subscripts $xl$, $zl$ denote respectively the components of the capillary force acting along the horizontal (x-) and vertical (z-) directions at the left ($l$) edge of the lamella underneath the macroscopic droplet contact line. Similarly, subscripts $xr$, $zr$ denote the same for the right edge ($r$) of the lamella. $ F_{zt}$ denotes the vertical capillary force acting on the top of lamella only at A-A$^\prime$ whereas  $F_{xr}$ and $F_{zr}$ are only applicable at C-C$^\prime$. The deflection profile $\delta(z)$ of a soft lamella due to the aforementioned capillary forces can be theoretically determined using variational principle as classically done for problems in linear elasticity\cite{Elastica_Audoly}. Specifically, the free energy functional for a deformed soft lamella (involving bending energy and the work done by the external capillary forces) is used to derive the corresponding Euler-Lagrange equation \cite{Elastica_Audoly}  as shown in Eq. (\ref{E:mm2}).
\begin{equation} \label{E:mm2}
B\theta''+F_x\cos\theta-F_z\sin\theta=0
\end{equation}
where $B=EI/(1-\nu^2)$ is the flexural rigidy for planar structures\cite{Landau_Lifshitz}, I is the moment of intertia about the bending axis (here y axis) and $\nu=0.5$\cite{PoissonRatioValue_1,PoissonRatioValue_2} is the Poisson's ratio for PDMS. Here, $F_x$ and $F_z$ are net horizontal and vertical forces acting on the lamella, i.e. resultants of $F_{xr}$, $F_{xl}$, and $F_{zr}$, $F_{zl}$, $F_{zt}$ respectively (see Eq. (\ref{Eqn_1}a) - Eq. (\ref{Eqn_1}e))  and $\theta'=\frac{d\theta}{ds}$ is the local curvature of deformed lamella with s as the arc length. So, the effects of $E$ and $\alpha$ are essentially encapsulated in $B$, and the effect of $\eta$ is implicitly represented by $F_x$ and $F_z$. Eq. (\ref{E:mm2}), along with the boundary conditions,

 \begin{gather}
s=0; \theta=0  \tag{\ref{E:mm2}a} \\
 s=L; \theta'=0 \tag{\ref{E:mm2}b}
 \end{gather}
 
is solved numerically using a shooting technique to evaluate the equilibrium shape of the deformed lamella. It is important to note here that the aforementioned solution procedure does not assume small deflection of the lamella as commonly done in the literature for quantifying the deformation of microstructures \cite{Manias,Rabodzey2008MechanicalFI}. For small deflection of the lamella an explicit expression for the shape of the deformed lamella can be written as Eq. (\ref{Eqn_3}) \cite{Gere_MOM}
\begin{equation} \label{Eqn_3}
\delta(z)=\frac{F_xL^3}{6B}\bigg(\frac{3z^2}{L^2}-\frac{z^3}{L^3}\bigg) 
\end{equation}
Furthermore, the values of $\psi$ and $\phi$ are also necessary for solving Eq. (\ref{E:mm2}). We estimate these values from  confocal images for different physical conditions, rather than using them as fitting parameters (see Sec. 7 in ESI). 
  
Fig. \ref{fig_5} shows comparisons between experimentally obtained (markers; see Sec. S4 in ESI for details) and theoretically evaluated (lines) deformation profiles of the lamella underneath the macroscopic droplet contact line at A-A$^\prime$ for different values of $\alpha$, $E$, and $\eta$. Note first that the theoretical deformation profile (dashed line) based on the small deflection assumption  given by Eq. (\ref{Eqn_3}) fails to describe the experimental deformation profile (Fig. \ref{fig_5}(a)(i) and (a)(ii)). However, the latter is described reasonably well by the general solution (solid line) obtained using Eq.  \ref{E:mm2}. Hence, it can be concluded that the small deflection assumption must not be assumed a priori to describe the wetting induced deformation of micro-structures, as often done in the literature\cite{Art_BatEl,Lemmon,Tan1484,Chandra}. A comparison of Fig. \ref{fig_5} (a)(i) (or (ii)) and Fig. \ref{fig_5} (b)(i) (or (ii)) shows that the smaller deformation of the lamella due to smaller value of $\alpha$, for identical value of $E$, is also reasonably addressed by the solution of Eq. (\ref{E:mm2}). Furthermore, at a particular value of $\eta$, the increasing deformation of the lamella underneath the macroscopic droplet contact line with reducing value of $E$ is also nicely captured (Fig. \ref{fig_5} (b)(i) and (ii)); this is mainly due to the reducing value of $B$ with decreasing $E$. Finally, Fig. \ref{fig_5}(a(i)) and Fig. \ref{fig_5}(a(ii)) show that the lamella deflection for $\eta=0$ and  $\eta=0.19$ at $z$ $\approx 30\mu m$ is $\delta \approx 3.28$ $\mu$m and $\delta \approx 2.87$ $\mu$m respectively. Hence, the deflection of lamella decreases with $\eta$. This mainly stems from the lowering of the horizontal component of the capillary force ($F_{xl}$ for A-A$^\prime$; Eq. (\ref{Eqn_1a})) triggered by the enhancement in $\psi$ due to the effect of Maxwell stress. It must be admitted here that the observable effect of $\eta$ on the deformation is indeed weak. However, provided the drop stays in the Cassie State, it is possible to achieve significant reduction in the deformation of the lamella with increasing $\eta$ (or increasing $\psi$) as predicted by the theoretical model (see Sec. S8 in ESI).
\begin{figure} [!h]
\centering
\includegraphics[width=6.91cm, height=12.16cm]{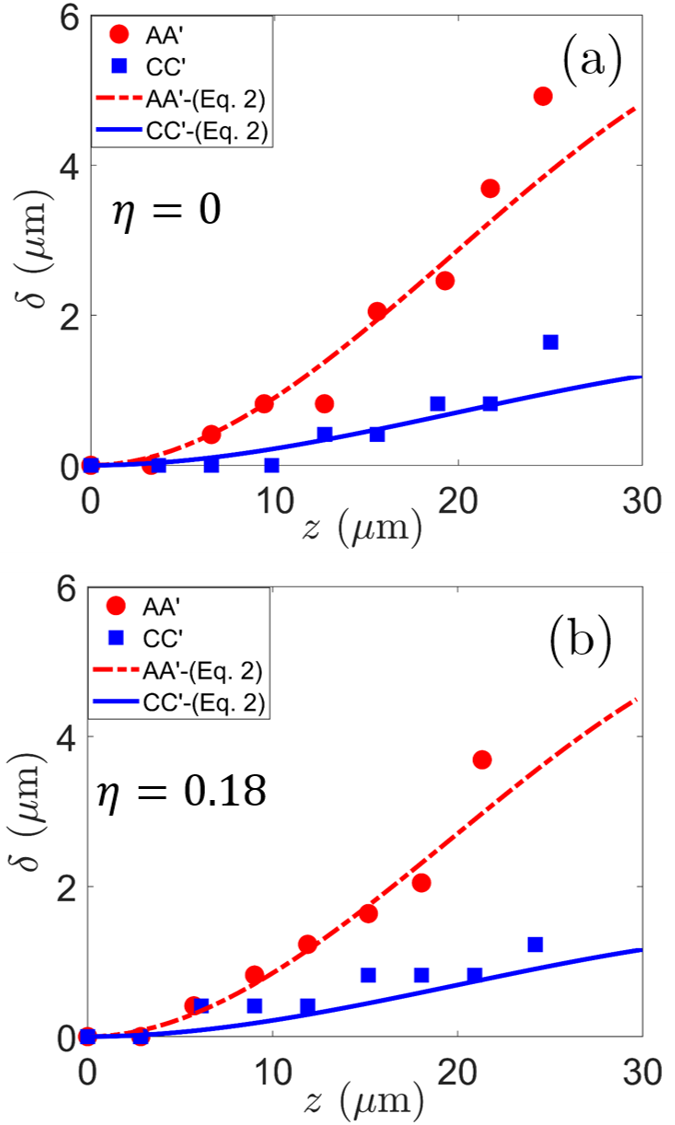}
\caption{\label{fig_6} Shapes of the deflected lamella (in the x-z plane) at the droplet contact line (markers: experimental data; lines: theory) in the A-A$^\prime$ plane (red) and in the C-C$^\prime$ plane (blue) for two different magnitudes of the electrowetting number $\eta$-- (a) $\eta=0$, (b) $\eta=0.18$. The aspect ratio $(\alpha)$ and the Young`s modulus $(E)$ of the deflected lamella shown here are 2.2 and 0.6 MPa respectively.}
\end{figure}
The solution of Eq. \ref{E:mm2} also nicely captures the smaller deformation of the lamella underneath the macroscopic droplet contact line at C-C$^\prime$ for different values of $\eta$ (Fig. \ref{fig_6}). The smaller deformation is because at C-C$^\prime$ there are capillary forces at both the top edges of the lamella (Fig. \ref{fig_2}(c)) which reduce the resultant horizontal capillary force on the lamella ($F_x=F_{xl}-F_{xr}$; (see Eq. \ref{Eqn_1})). Furthermore, at C-C$^\prime$, the maximum deformation of the lamella is also slightly less for $\eta=0.18$ than for $\eta=0$. Based on the aforementioned discussion, it can be now intuitively concluded that the lamellae underneath the droplet but away from the macroscopic contact line do not deform because the horizontal components of the capillary forces acting on the two top edges of each are almost equal in magnitude but opposite in direction thereby resulting in zero net horizontal force.
 
\section{Conclusion}
We studied the microscopic deformation of elastic lamellae constituting a super-hydrophobic substrate during electrowetting of a sessile drop using confocal microscopy. We provide a detailed description of the local deformation of the lamella, which turns out to be mainly controlled by the net horizontal component of the surface tension force exerted by the liquid-vapor interface. Since the orientation of the free liquid surface and the pinning conditions of the three phase contact line varies along the macroscopic edge of the drop, the local deformation of the lamella also varies and is maximum in regions where the liquid-vapor interface is essentially only pinned to one edge of a lamella but not to the other. The role of electrowetting is indirect: by moving the contact line and deforming the liquid surface, it controls the magnitudes of the net capillary forces and the direction in which these pull. A direct influence of the Maxwell stress on the deformation of the lamella, however, could not be detected - and is in fact expected to be very small. Analyzing the shape and maximum deflection of the lamella and comparing to theory, we find that the frequently used small deflection model fails to provide a quantitative description of the experimental results. A full non-linear system of equation and boundary conditions (albeit derived using linear elasticity theory) that requires a numerical solution, however, describes the experiments very well. This conclusion may also apply to other experiments that were previously analysed in terms of a linear small deflection model.

\section*{Acknowledgement}
AAD and RD sincerely thank Jacco Snoeijer for his wonderful course on `Fluids and Elasticity' in which the basic concepts of linear elasticity and the application of variational principle were introduced. AAD and FM also thank Hans-Jurgen Butt for hosting the former at MPI Mainz for a summer internship during the course of which the basics of the fabrication process used in this work were explained and demonstrated. The authors also  thank Daniel Wijnperle for his immense help with the fabrication process.\\

%%%REFERENCES%%%

\bibliography{ms_bib} %You need to replace "rsc" on this line with the name of your .bib 

\bibliographystyle{rsc} %the RSC's .bst file

\end{document}